# The Tausk Controversy on the Foundations of Quantum Mechanics: Physics, Philosophy, and Politics[#]

Osvaldo Pessoa, Jr., Olival Freire, Jr., and Alexis De Greiff

Abstract: In 1966 the Brazilian physicist Klaus Tausk (b. 1927) circulated a preprint from the International Centre for Theoretical Physics in Trieste, Italy, criticizing Adriana Daneri, Angelo Loinger, and Giovanni Maria Prosperi's theory of 1962 on the measurement problem in quantum mechanics. A heated controversy ensued between two opposing camps within the orthodox interpretation of quantum theory, represented by Léon Rosenfeld and Eugene P. Wigner. The controversy went well beyond the strictly scientific issues, however, reflecting philosophical and political commitments within context of the Cold War, the relationship between science in developed and Third world countries, the importance of social skills, and personal idiosyncrasies.

[#] This paper is now accepted to be published in the journal "Physics in Perspective" [PiP]. After its publication in PiP, we can post its full text on the web.